# First-Principles Investigation of the Pressure Dependent Physical Properties of Intermetallic Kagome ZrRe$_2$

Mst. Irin Naher[1*], A. F. M. Yusuf Haider[1], Dholon Kumar Paul[1], Md Lutfor Rahman[1], Firoze H. Haque[1], Saleh Hasan Naqib[2*]

[1]Department of Mathematics & Physical Sciences, Brac University, Dhaka1212, Bangladesh
[2]Department of Physics, University of Rajshahi, Rajshahi 6205, Bangladesh
*Corresponding authors; Emails: irinnaher.phy@gmail.com & salehnaqib@yahoo.com

**Abstract**
Intermetallic compounds hosting Kagome lattices have drawn great interest as a vibrant emerging field in condensed matter physics. We present a first-principles density functional theory (DFT) investigation of the pressure-dependent structural, electronic, mechanical, thermophysical, vibrational, and optical properties of the intermetallic Kagome compound ZrRe$_2$. The calculated ground-state structural parameters are in excellent agreement with available experimental results. The estimated structural parameters, elastic constants, and phonon dispersion confirm the structural, chemical, mechanical, and dynamical stability of ZrRe$_2$ up to 25 GPa. The Kagome feature in the material has been identified from the electronic band structure for the first time. ZrRe$_2$ exhibits topological feature at 0 GPa, which vanishes under 25 GPa. Fermi surface (FS) analysis predicts that ZrRe$_2$ could potentially host a charge density wave (CDW) phase. The electronic and optical studies confirmed its metallic nature. The estimated Pugh's ratio, Poisson's ratio, and machinability index indicate that the compound remains ductile and exhibits excellent machinability within the studied pressure range. The Debye temperature and phonon thermal conductivity are moderate, while the melting point is relatively high. Furthermore, ZrRe$_2$ possesses moderate electron-phonon coupling, which weakens under pressure as the phonon modes harden. Consequently, the superconducting transition temperature decreases with increasing pressure. The compound also shows potential as a promising TBC (thermal barrier coating) material. Most of the properties studied and analyses performed in this paper are novel in nature.

## 1. Introduction

Intermetallic Kagome systems have attracted major theoretical and experimental interest in the field of condensed matter physics and quantum physics due to their unique combination of intermetallic stability and Kagome-derived electronic properties [1-5]. However, only about 18% of intermetallics are reported to contain Kagome layers [1]. Intermetallics themselves are among the oldest known classes of materials in condensed matter physics, yet they continue to attract attention for their remarkable physical and mechanical properties. Large number of intermetallic compounds exhibit attractive combination of physical and mechanical properties, including high melting point, low diffusivity, high creep resistance, good oxidation or corrosion resistance, low densities, excellent high-temperature strength, and in some cases, ductility [6-7]. Beyond these well-established features, a particularly exciting direction has emerged with the discovery of intermetallics hosting Kagome lattices.

A Kagome lattice, consisting of a two-dimensional (2D) network of corner-sharing triangles, has drawn significant attention since its introduction to in quantum physics in 1951. Stacked Kagome layers can host a variety of emergent phenomena, including geometrically frustrated magnetism, nontrivial band topology, correlated electron orders, charge/spin density waves (CDW/SDW), pair density wave, quantum spin liquids (QSL), hall effect, Chern magnetism, and unconventional superconductivity [8-12]. Moreover, their electronic band structures naturally accommodate Dirac points, flat bands, and van Hove singularities, thereby promoting non-trivial topological states and robust electronic orders [2].

When Kagome layers are incorporated into intermetallic compounds, they give rise to exotic electronic and topological features, enabling multifunctional materials for next-generation energy, spintronics, and quantum technologies [2]. These properties also provide pathways for tuning exotic states through external magnetic fields, high-pressure, and high-temperature engineering [4]. Hence, understanding the influence of pressure on such materials is crucial for the design of robust spintronic and quantum devices.

Although $ZrRe_2$ was first synthesized in 1942, only a few studies have explored its properties so far [13-17]. It exhibits high melting temperature of 2750±50° K and a microhardness of 1200 kg/mm$^2$. Recently, Y. Yu et al. demonstrated the existence of Kagome layers in intermetallic $ZrRe_2$, with weakly-coupled Type-II superconductivity with $T_c$ ~ 6.1 K, and reported lower and upper critical fields of 6.27 mT and 12.77 T, respectively [17]. $ZrRe_2$ is metallic, with density of states at the Fermi level $N(E_F) = 2.8114 \ eV^{-1}.f.u.^{-1}$, Normalized specific heat change $\Delta C_e / \gamma T_c =$ 1.24, Debye temperature $\Theta_D = 301$ K, and electron-phonon coupling strength $\lambda_{ep} = 0.69$ [17]. First-principles studies have investigated only it's mechanical properties and electronic structure (band, DOS, and Fermi surface) with and without spin orbit coupling (SOC) of $ZrRe_2$ at ambient pressure so far [15,17].

To the best of our knowledge, several important physical properties and their pressure dependency, including electronic (band, DOS, charge density difference, Fermi surface with and without SOC), mechanical, phonon, thermodynamic, and optical, of $ZrRe_2$ have not yet been systematically explored. In this work, we have performed a set of first-principles calculations and explored the pressure-dependent electronic (with and without SOC), mechanical, thermodynamic, phonon, and optical properties of newly reported intermetallic Kagome $ZrRe_2$ for the very first time. Pressure-dependent investigations provide insights into the stability and reversibility of exotic states under strain or pressure. Therefore, through a comprehensive analysis of these properties in relation to the existing features of $ZrRe_2$, we seek to bridge current knowledge gaps, lay the groundwork for future experimental studies, and unlock its potential for industrial applications.

The rest of this manuscript has been organized as follows: Section 2 provides a brief description of the computational scheme. Section 3 presents the results and discussion. Finally, Section 4 contains conclusions.

## 2. Computational scheme

We performed DFT simulation within the generalized gradient approximation (GGA) [18] with the Perdew-Burke-Ernzerhof for solids (PBEsol) scheme for the exchange-correlation potential as employed in the CASTEP code [19-20]. The Zr-$4s^2$ $4p^6$ $4d^2$ $5s^2$ electrons and Re-$5s^2$ $5p^6$ $5d^5$ $6s^2$ electrons are treated as valance states. The Vanderbilt-type ultrasoft pseudopotential was applied to represent the interaction the interaction between the valence electrons and ion cores [21]. To ensure the convergence of the structure and energies, the plane-wave basis set cut-off energy and Monkhorst-Pack grid [22] in the irreducible Brillouin zone (BZ) were set to 500 eV and 7×7×5, respectively. A total energy convergence of $0.5\times10^{-5}$ eV/atom, maximum ionic force of 0.01 eV/Å, maximum ionic displacement of $0.5\times10^{-3}$ Å, maximum stress of 0.02 GPa, and smearing width of 0.1 eV were used during the optimization process. A denser $k$-point mesh of 22×22×12 was used for precise determination of the Fermi surface. The second-order elastic stiffness constants ($C_{ij}$) were determined using the stress–strain relation [23]. ELATE code [24] was used to generate a 3D elastic anisotropy contour. The phonon dispersion and phonon density of states were investigated using the density functional perturbation theory (DFPT) as implemented in the CASTEP code. The optical features were executed by applying the Kramers–Kronig relations [25].

## 3. Results and discussion
### 3.1. Structural properties

Fig. 1 presents the crystal structure of the Kagome metal ZrRe$_2$. The two stacked Kagome nets of Re1 atoms along the $c$ direction is interleaved with triangular layers, shown in Fig. 1b, along with an indication of a charge density wave (CDW) in the ab-plane. A breathing Kagome lattice emerges, characterized by two distinct Re1–Re1 bond lengths of 2.70 Å and 2.56 Å. The close-packed Zr-Re2 layer is capped above and below by the Kagome nets. ZrRe$_2$ crystallizes into hexagonal structure with a formula unit Z = 4 and space group P6$_3$/mmc (No. 194). The Wyckoff positions of the constituent atoms in the unit cell are: Zr (0.333, 0.667, 0.061), Re1 (0, 0, 0), and Re2 (0.829, 0.658, 1/4). Table 1 summarizes the calculated equilibrium lattice constants ($a$, $c$), $c/a$ ratio, volume (V), and bond length ($d$) of ZrRe$_2$ at 0 GPa and 0 K, obtained without SOC and with SOC. Available experimental and theoretical data are presented for comparison. Excellent agreement is observed between our estimated structural parameters and the experimental data [17,26,27]. The optimized structure exhibits closer correspondence to experiment than earlier theoretical studies [15], implying higher accuracy in our results. The effect of SOC does not result in any appreciable changes in structural parameters. The normalized structural parameters (without SOC) as a function of pressure are displayed in Fig. 2a. The absence of any abrupt changes in the structural parameters indicating structural stability of ZrRe$_2$ within the studied pressure range. At higher pressure, the material starts compressing more along the $a$-axis compared to the $c$-axis.

Although several studies have investigated ZrRe$_2$, the stability of the material has not yet been addressed theoretically, even after its successful synthesis. Since stability is crucial for understanding potential applications, we have investigated the pressure-dependent chemical stability of ZrRe$_2$ between 0-25 GPa (see Fig. 2b).

The chemical stability of ZrRe$_2$ is evaluated using the following formula:

$$E_f^{ZrRe_2} = \frac{E_{total}^{ZrRe_2} - (xE_{solid}^{Zr} + yE_{solid}^{Re})}{x + y} \quad (1)$$

here, x = 4 and y = 8 are the number of Zr and Re atoms, respectively, in the unit cell of ZrRe$_2$. $E_{total}^{ZrRe_2}$ is the total energies of the ZrRe$_2$ compound, while $E_{solid}^{Zr}$ and $E_{solid}^{Re}$ refer to the total energies of elemental Zr and Re in their stable solid-state crystal structures. Formation enthalpy ($\Delta H_f$) of the material at ambient pressure is estimated using the open quantum materials database (OQMD). The calculated $E_f$ and $\Delta H_f$ of ZrRe$_2$ are found to be -0.663 eV/atom and -0.361 eV/atom at 0 GPa and 25 GPa, respectively. The negative values for both pressures matrices indicate the chemical stability of ZrRe$_2$ at ambient pressure and under pressure (see Fig. 2b).

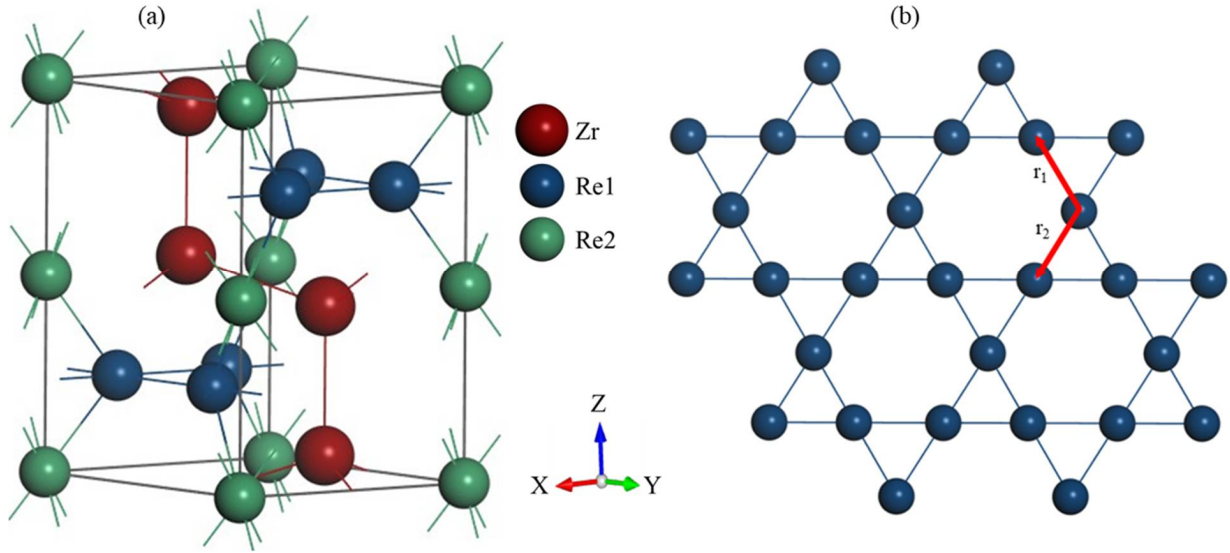

**Fig. 1.** (a) Schematic 3D Crystal structure of ZrRe$_2$ and (b) Breathing Kagome lattice formed by Re1 atoms ($r_1 \neq r_2$).

**Table 1:** Lattice parameter (*a* and *c*), *c/a* ratio, volume (*V*), and bond length (*d*) of ZrRe$_2$ without SOC and with SOC at 0 GPa and 0 K.

| | a (Å) | c (Å) | c/a | V (Å$^3$) | Bond length (Å) | | | Ref. |
|---|---|---|---|---|---|---|---|---|
| | | | | | $d_{Re-Re}$ | $d_{Zr-Re}$ | $d_{Zr-Zr}$ | |
| Without SOC | 5.258 | 8.627 | 1.641 | 206.551 | 2.558; 2.661 | 3.080; 3.095 | 3.211 | This |
| With SOC | 5.259 | 8.632 | 1.641 | 206.652 | - | - | - | |

| | | | | | | | |
|---|---|---|---|---|---|---|---|
| 5.268 | 8.630 | 1.638 | - | 2.58; 2.68 | - | - | [17][Expt.] |
| 5.262 | 8.593 | 1.633 | 206.052 | - | - | - | [26][Expt.] |
| 5.251 | 8.576 | 1.633 | 204.786 | - | - | - | [27][Expt.] |
| 5.327 | 8.709 | 1.635 | 214.038 | - | - | - | [15][Theo.] |

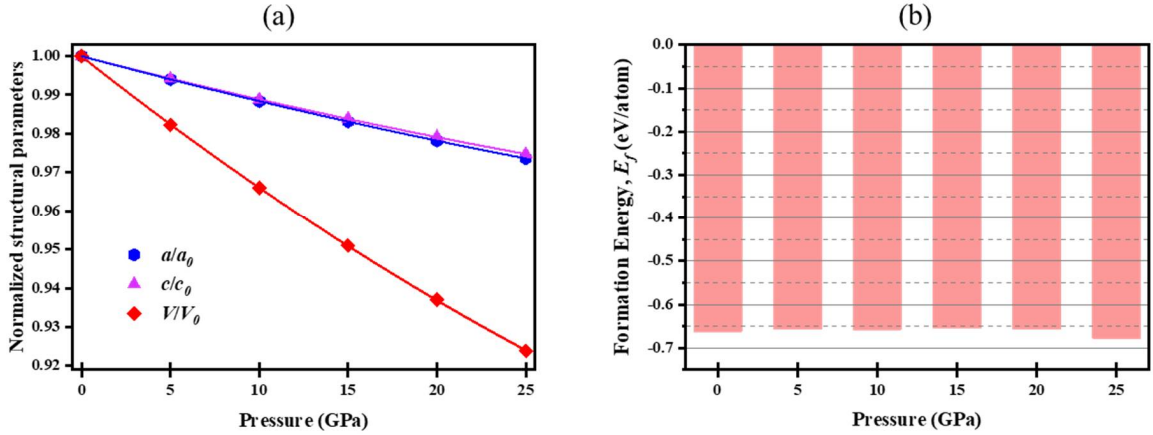

**Fig. 2.** Pressure dependence of (a) normalized structural parameters and (b) formation energy of ZrRe$_2$ without SOC.

## 3.2. Electronic properties

To further understand the electronic feature of ZrRe$_2$, we performed electronic band structure, electronic energy density of states (DOS), electron density difference (EDD), and Fermi surface (FS) calculations (Fig. 3-7).

### 3.2.1. Band structure and density of states

The electronic band structure of ZrRe$_2$ without and with SOC effect at 0 and 25 GPa is depicted in Fig. 3. The horizontal dashed line in band represents Fermi level ($E_F$). Band structure of ZrRe$_2$ is intricate with large number of bands crossing the $E_F$ which indicates the high metallicity of the material. Band degeneracy at several points in the Brillouin zone is a result of high symmetry [2]. Contrary to previous findings [17], which reported no Dirac cones, van Hove singularities, or flat bands near the Fermi level under either SOC or non-SOC conditions, our band structure analysis reveals clear topological and Kagome features. At 0 GPa, there is a Dirac cone at $\Gamma$ near $E_F$ (inset in Fig. 3) without SOC and the gap appears with SOC. The Dirac point moves downward with pressure and disappears around a critical pressure $P_c$ ~ 12 GPa. After critical pressure, the gap increases steadily under pressure. Flat bands centered at the $\Gamma$-point very close to the Fermi level are signatures of strong electronic correlations and possible van Hove singularities in the electronic energy density of states.

Fig. 4 displays the total and partial density of states (DOS) of ZrRe$_2$ at different hydrostatic pressures. The vertical broken line at zero energy represents the Fermi level. The high total DOS at the $E_F$ demonstrates the metallic nature of the compound. The value of $N(E_F)$ at 0 GPa is found to be 11.59 electrons.eV$^{-1}$.cell$^{-1}$ (2.90 electrons.eV$^{-1}$.f.u.$^{-1}$), showing good consistency with earlier theoretical reports [17]. Strong hybridization between Zr 4$d$ and Re 5$d$ around $E_F$ indicating of the formation of strong Zr-Re covalent bonding. The electronic bands and DOS of ZrRe$_2$ near the $E_F$ are mainly dominated by Re 5$d$ electronic states. In contrast, Zr contributes only weakly to the valence region, acting mainly as an electron donor. The high $N(E_F)$ in Kagome materials is reported to originate from the weakly dispersive bands of the Kagome lattice, which become broadened due to interlayer interactions [5]. The superconductivity of ZrRe$_2$ might associated with either the Zr sublattice, the Re sublattice, or both, as each contributes to the density of states around the Fermi level. Kagome-like bands bear a strong contribution of Re1 5$d$ states around the $E_F$, while the Re2 5$d$ mostly contributes to a set of weakly dispersive bands around − 2 to − 3 eV below the $E_F$ (not shown here). The $N(E_F)$ decreases slightly under 25 GPa. The peak broadening in the TDOS observed at 25 GPa can be attributed to the reduction in inter-atomic distances, which enhances the overlap of atomic states.

Inclusion of spin–orbit coupling (SOC) slightly decreases the value of DOS near the $E_F$. Interestingly, it also lifts a sharp peak in TDOS (see Fig. 5) and band degeneracy (not shown here) around 6 eV upon SOC inclusion. This sharp peak is likely associated with the Zr $d$ states, consistent with earlier reports in some cubic AB$_2$ type intermetallic compounds in the literature [28]. To further examine possible spin polarization, we also calculated the electronic band structure with intrinsic spin-polarized effects (not shown here). The nearly identical band structures obtained with and without spin polarization confirm that the material is nonmagnetic.

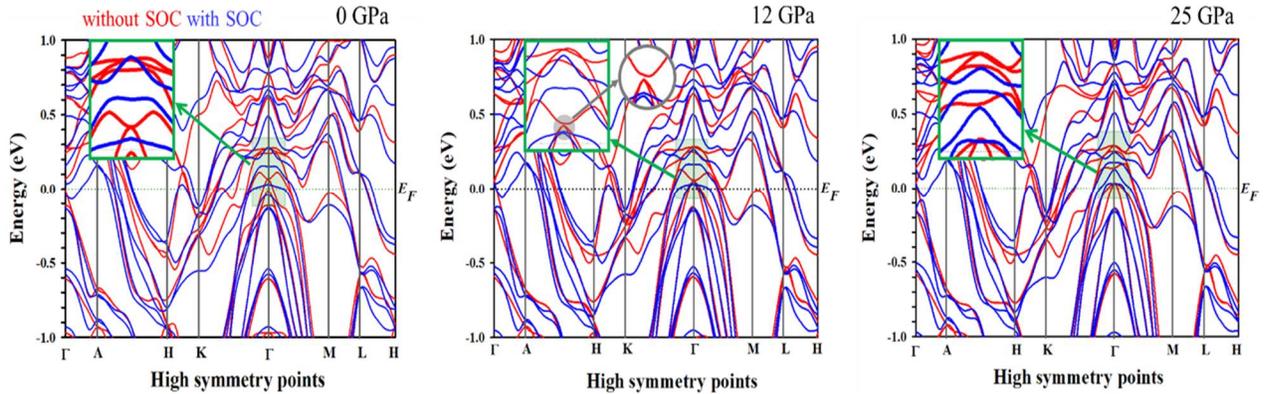

**Fig. 3.** Calculated electronic band structure of ZrRe$_2$ without and with SOC at 0, 12, and 25 GPa.

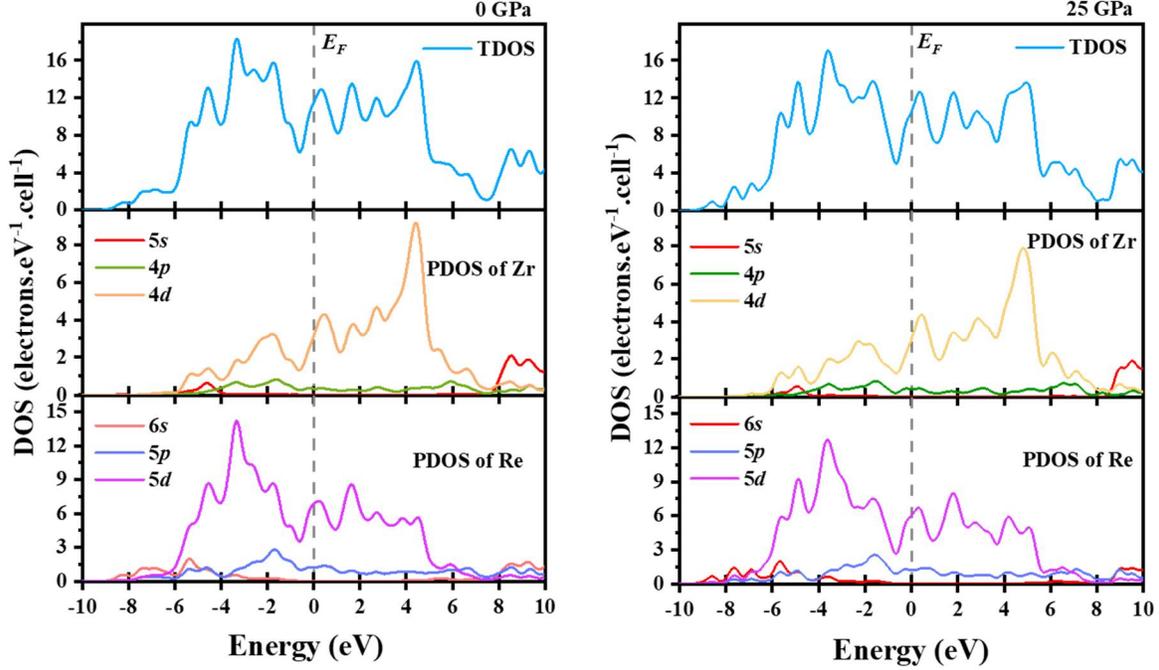

**Fig. 4.** The electronic total and partial density of states of ZrRe$_2$ without SOC at 0 and 25 GPa.

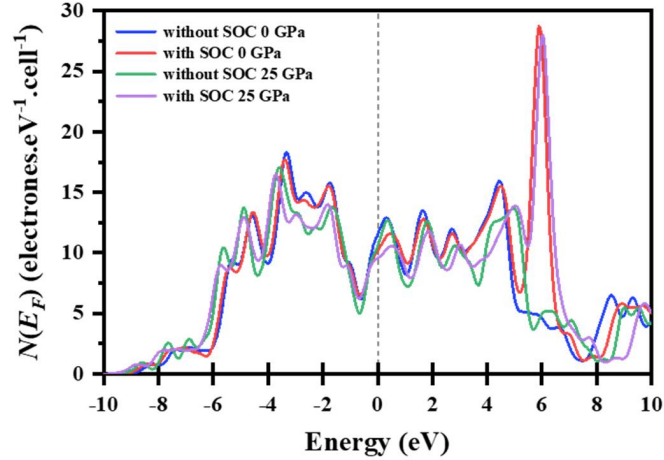

**Fig. 5.** Total density of states of ZrRe$_2$ without and with SOC at 0 and 25 GPa.

*3.2.2. Electron density difference*

The electron density difference (EDD) is very important to elucidate the bonding nature and electron distribution of each element in the material. In Fig. 6, the calculated EDD of the compound at 0 GPa without SOC is visualized. The color bar on the right indicates the EDD values in e/Å$^3$, where red and blue represent regions of maximum and minimum electron density, respectively. The yellow-greenish region surrounding Zr atoms indicating moderate electron depletion and suggesting that Zr acts as an electron donor. Positive EDD values between Re atoms exhibits strong Re-Re covalent bonding. The yellowish contours between Zr and Re atoms suggest minimal charge transfer and weaker bonding relative to Re-Re. These observations are in excellent agreement with results discussed in Section 3.2.1. Anisotropy in charge distribution is also evident in different planes. Furthermore, a significant directional dependency in the charge distribution is evident, signifying the presence of covalent bonding. The effect of pressure appears negligible (not shown here).

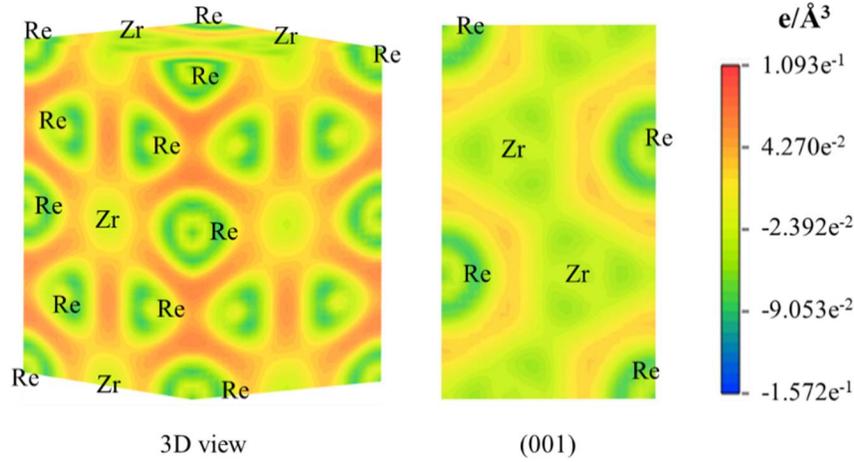

**Fig. 6.** The electron density difference maps for ZrRe$_2$ without SOC at 0 GPa.

*3.2.3. Fermi surface topology*

Electrical, magnetic, thermal, and optical nature of metals and semimetals are correlated with their Fermi surface (FS) topology. The FS is the constant energy surface and the velocity of the charge carriers is normal to the surface. The FS of intermetallic Kagome ZrRe$_2$, constructed from the relevant band structure without SOC at 0 and 25 GPa, are depicted in Figs. 7-8. Both electron- and hole-like sheets form the FS. At ambient pressure, electron-like cylindrical FSs centered around the $\Gamma$-point (along the $\Gamma$-A path) arise from the bands 81, 82, 83, and 84, implying a strong two-dimensionality of the electronic properties. For bands 86 and 87, a tiny hole-like sheet appears around the K-point. The constructed FS topology at 0 GPa matches the literature [17]. The FS of ZrRe$_2$ under 25 GPa are quite similar to those at ambient pressure, except for the appearance of one tiny electron-like extra sheet (band 80) around the $\Gamma$-point. The cylindrical Fermi sheets centered on the $\Gamma$-point has portions that are almost parallel to each other leading to Fermi surface nesting. This feature is well known to drive CDWs formation in strongly correlated electronic systems [2].

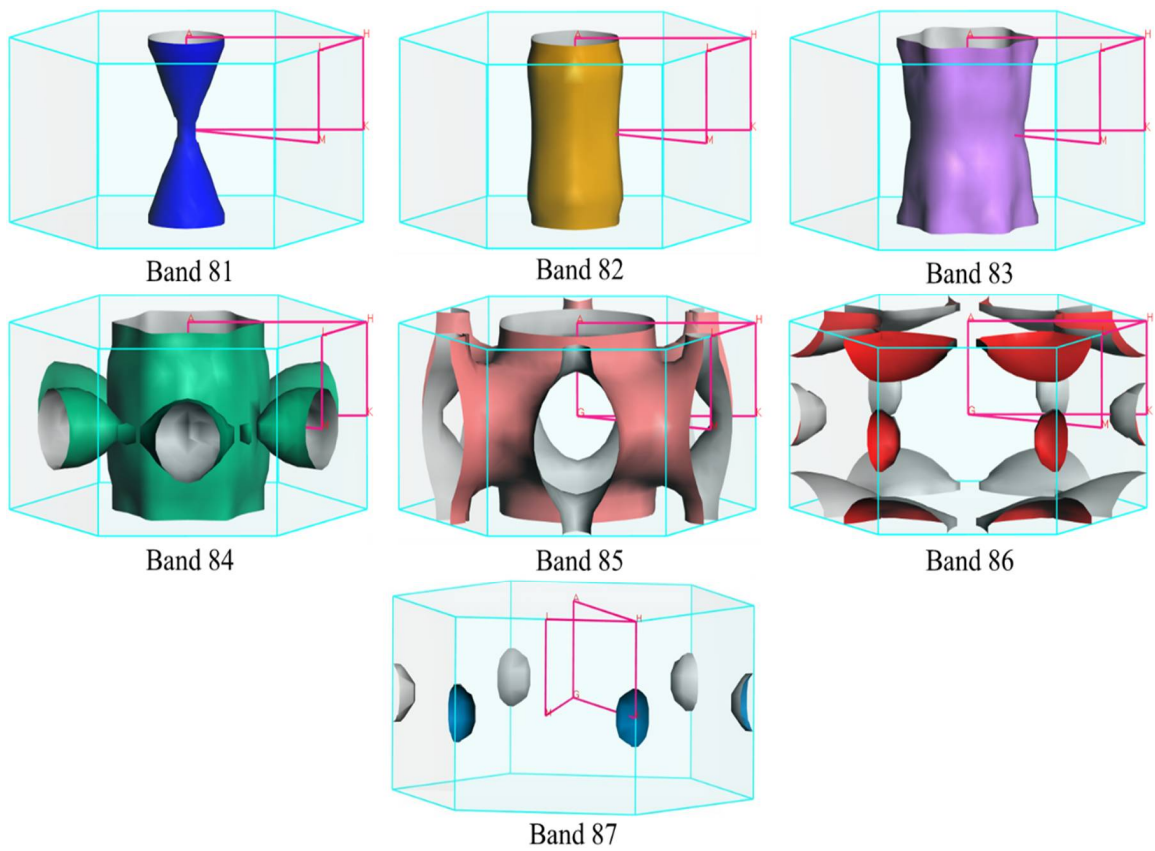

**Fig. 7.** Fermi surface topologies of ZrRe$_2$ without SOC at 0 GPa.

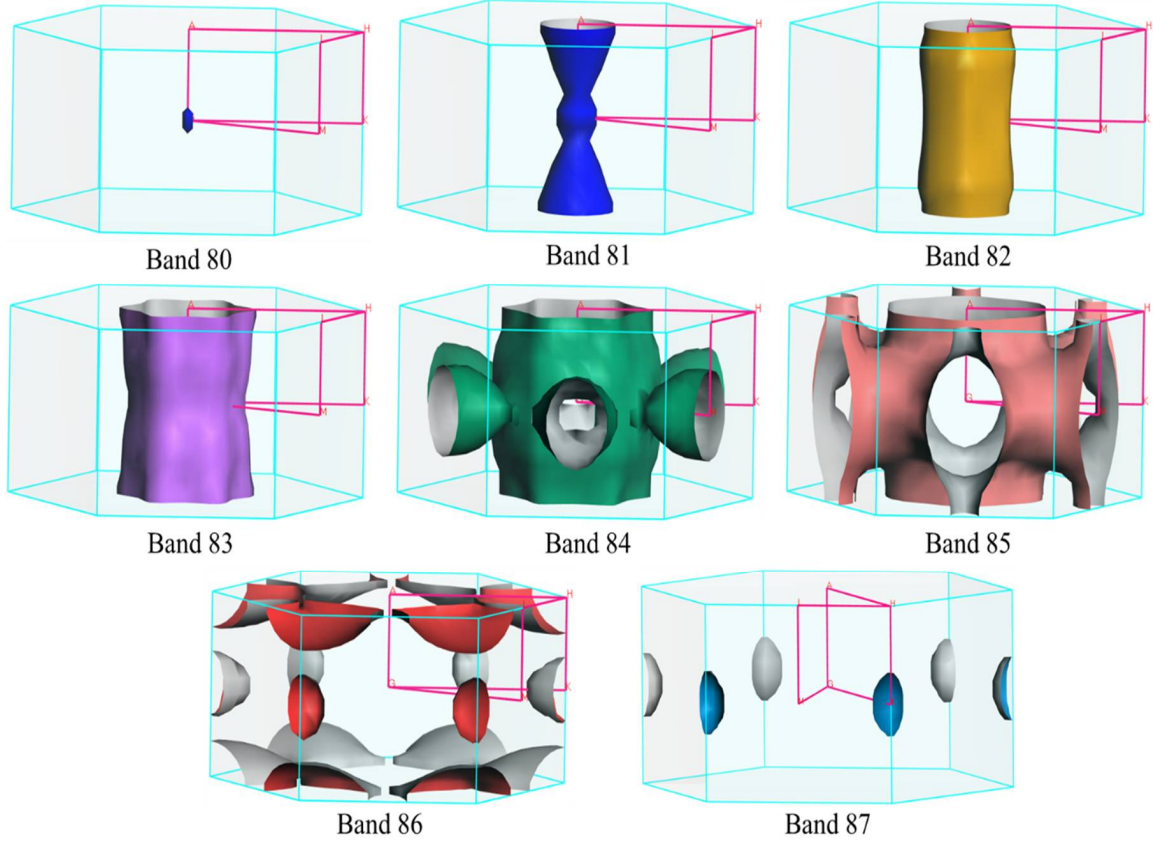

**Fig. 8.** Fermi surface topologies of ZrRe$_2$ without SOC at 25 GPa.

### 3.3. Mechanical and elastic features

Engineering materials often possess complex geometries and exhibit considerable variation in their dimensional attributes, resulting in highly non-uniform stress distributions within the component. Consequently, a thorough investigation of the elastic stiffness constants is fundamental for elucidating the intrinsic mechanical behavior and elastic limits of materials under external stress, which in turn is indispensable for accurately evaluating their potential in advanced engineering applications. Throughout the evolution of scientific research, stiffness constants have served as fundamental parameters in numerous fields, encompassing materials science, condensed matter physics, engineering, chemistry, and geophysics [2]. The reader is referenced to Ref. [2,29-31] for an in-depth discussion and detailed insights into the elastic property relationships. As the studied material is hexagonal, there will be five independent elastic stiffness constants: $C_{11}$, $C_{33}$, $C_{44}$, $C_{12}$, and $C_{13}$. The mechanical stability of ZrRe$_2$ is confirmed when the examined elastic constants satisfy the Born's stability criteria [32]: $C_{11} - |C_{12}| > 0$, $(C_{11} + C_{12})C_{33} - 2C_{13}^2 > 0$, $C_{44} > 0$. Furthermore, the mechanical stability criteria for hexagonal crystal under pressure (P) are expressed as [32]:

$$\tilde{C}_{44} > 0, \tilde{C}_{11} > |\tilde{C}_{12}|, \tilde{C}_{33}(\tilde{C}_{11} + \tilde{C}_{12}) > 2\tilde{C}_{13}^2 \qquad (2)$$

where, $\tilde{C}_{ii} = C_{ii} - P (i = 1, 3)$ and $\tilde{C}_{1i} = C_{1i} + P (i = 2, 3)$

All the calculated elastic stiffness constants ($C_{ij}$) of ZrRe₂ satisfy these stability criteria within the pressure range of 0–25 GPa (Fig. 9a). Therefore, ZrRe₂ can be regarded as mechanically stable up to 25 GPa. The independent elastic constants for ZrRe₂ at ambient pressure align well with the literature [15]. Although the material shows a general enhancement in stiffness with applied pressure, the distinct slopes of the elastic constants indicate unequal pressure sensitivities among them. The constants $C_{11}$ and $C_{33}$ define the atomic bonding strength and resistance to linear compression along the [100] and [001] directions, respectively. In contrast, $C_{12}$ and $C_{13}$ describe the resistance to shape deformation along the [010] and [001] directions. The elastic constant $C_{44}$ corresponds to the resistance to shear deformation under tangential stress applied to the (100) plane in the [010] direction [31]. $C_{44}$ is directly associated with hardness and valence electron density of a material, whereas it exhibits an inverse correlation with its machinability. The values of $C_{11}$ and $C_{33}$ are relatively higher. Under pressure, the material is more compressible along [100] than along [001]. Although, $C_{44}$ increases slightly but remains the smallest in magnitude, implying that shear resistance is comparatively weak and insignificant pressure dependence.

The bulk ($B$), shear ($G$), and Young's ($E$) moduli were estimated using the second-order elastic stiffness constants ($C_{ij}$) under the Voight-Reuss-Hill scheme [33-35]. The corresponding formulae are summarized elsewhere [30,36]. Fig. 9b illustrates the variation of $B$, $G$, and $E$ of ZrRe₂ with pressure. All the calculated values at ambient pressure are in good agreement with the literature [15]. The bulk modulus ($B$) reflects the bonding strength and resistance to volume deformation whereas the shear modulus ($G$) characterizes the degree of plasticity and the resistance to shear or transverse distortion. The relatively smaller value of $G$ compared with $B$ for ZrRe₂ implies that its mechanical strength is primarily governed by shape deformation mechanisms up to 25 GPa. The Young's modulus ($E$) represents the overall stiffness of the material and is also related to its thermal shock resistance. A high $E$ value of ZrRe₂ indicates substantial stiffness and good thermal shock resistance, making it a promising candidate for thermal barrier coating (TBC) applications. All three moduli increase with increasing pressure, with the pressure derivative of the bulk modulus being the highest, suggesting that ZrRe₂ becomes progressively less compressible and structurally stiffer under compression.

Pugh's ratio ($B/G$) is a widely used parameter by metallurgists to quantify the brittleness or ductility of a material. The materials are considered ductile when $B/G > 1.75$; otherwise, they are classified as brittle. A higher (or lower) $B/G$ value indicates greater ductility (brittleness) of the material. As shown in Fig. 9c, ZrRe₂ is intrinsically ductile up to 25 GPa and ductility increases under pressure. The brittleness and ductility of materials can also be confirmed from another parameter called Poisson's ratio ($v$), defined as: $v = (3B - 2G)/2(3B + G)$ [29]. The brittle and ductile materials satisfy condition $v \leq 0.26$ and $v \geq 0.26$, respectively [37]. The estimated $v$ value also (like Pugh's ratio) suggests ductility of the compound up to 25 GPa (Fig. 9c). Overall, both parameters are in good agreement with previously reported results in the literature [15]. The finding that ZrRe₂ exhibits ductility is particularly noteworthy, as intermetallic compounds are typically brittle. This property could make ZrRe₂ a promising candidate for mechanically stable engineering or quantum devices.

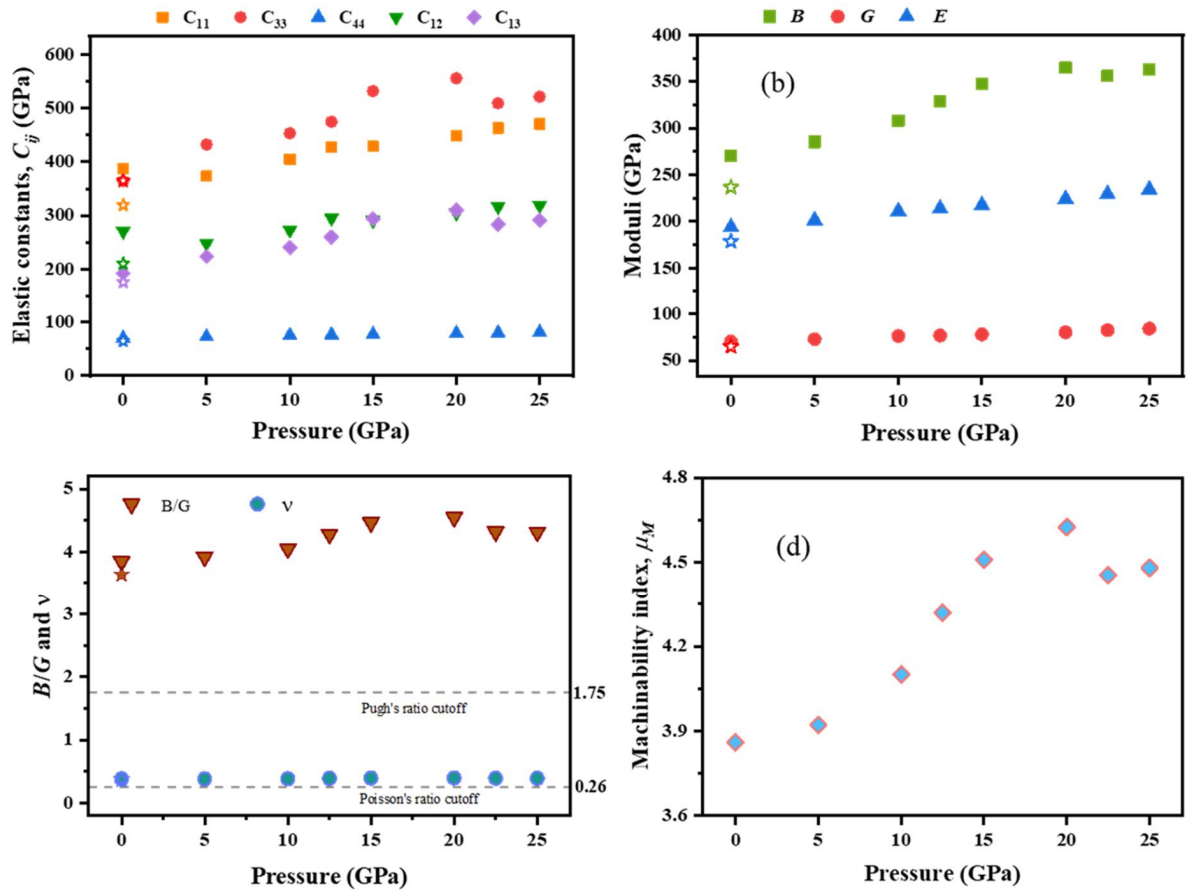

**Fig. 9.** The pressure dependent (a) elastic constants, (b) Moduli, (c) Pugh's and Poisson's ratio, and (d) machinability index of ZrRe$_2$. The star symbols in each plot represent earlier theoretical values corresponding to the respective color.

Another vital parameter for selecting engineering materials is machinability. The machinability of engineering materials plays a crucial role in optimizing processing and machining operations, as it directly influences cutting conditions, tool wear, and overall efficiency [2,30]. The machinability index ($\mu_M$) defined as $\mu_M = B/C_{44}$ and proposed by Sun et al. [38], provides a useful measure of this property. The estimated $\mu_M$ values of ZrRe$_2$ up to 25 GPa indicates that the material is expected to exhibit excellent machinability (Fig. 9d). Materials with high $\mu_M$ typically exhibit enhanced lubricating and adaptable behavior, along with lower feed force, and higher plastic strain and surface roughness values [39]. Comparing the elevated $\mu_M$ value of ZrRe$_2$ with that of other well-known lubricating materials suggests that it possesses outstanding lubricating characteristics [2].

The mechanical anisotropy of ZrRe$_2$ as a function of pressure is studied using universal anisotropy index ($A^U$) [40], defined as:

$$A^U = 5\frac{G_V}{G_R} + \frac{B_V}{B_R} - 6 \geq 0 \tag{3}$$

where, $G_V$ and $G_R$ are the Voigt and Reuss shear moduli, and $B_V$ and $B_R$ are the Voigt and Reuss bulk moduli, respectively. The variation of $A^U$ with pressure is shown in Fig. 10. $A^U = 0$ for isotropic crystals and becomes a finite value ($> 0$) in the case of elastic anisotropy. Positive values of $A^U$ for ZrRe$_2$ confirm their significant anisotropic nature under 0-25 GPa. However, the anisotropy decreases markedly at 5 GPa and after that pressure effect is minimal under studied pressure range.

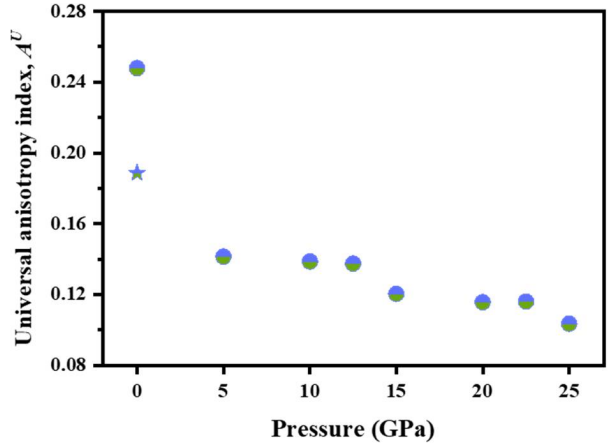

**Fig. 10.** Pressure dependent universal anisotropy index for ZrRe$_2$. The star symbol indicates the theoretical reference.

Investigating the directional anisotropy of mechanical parameters is crucial for material selection and design planning in applications to achieve optimal performance. In addition to estimating overall mechanical anisotropy using $A^U$, we also studied the anisotropy in Young's modulus ($E$), shear modulus ($G$), linear compressibility ($\beta$), and Poisson's ratio ($v$) by constructing their three-dimensional (3D) graphical visualizations. The 3D representations of $E$, $G$, $\beta$, and $v$ for the compound at 0 and 25 GPa are shown in Fig. 11. An ideal spherical shape represents isotropic behavior. At ambient pressure, the linear compressibility exhibits least anisotropy and becomes almost isotropic under 25 GPa. All the four parameters exhibit isotropic behavior around $z$-axis at both pressures. Moreover, Poisson's ratio shows the highest degree of anisotropy. Therefore, the anisotropy in the material follows the order: $\beta < E < G < v$. The overall anisotropy of all four parameters decreases at 25 GPa (see Fig. 11 and Table 2), which is also consistent with the trend in the mechanical anisotropy (see Fig. 10).

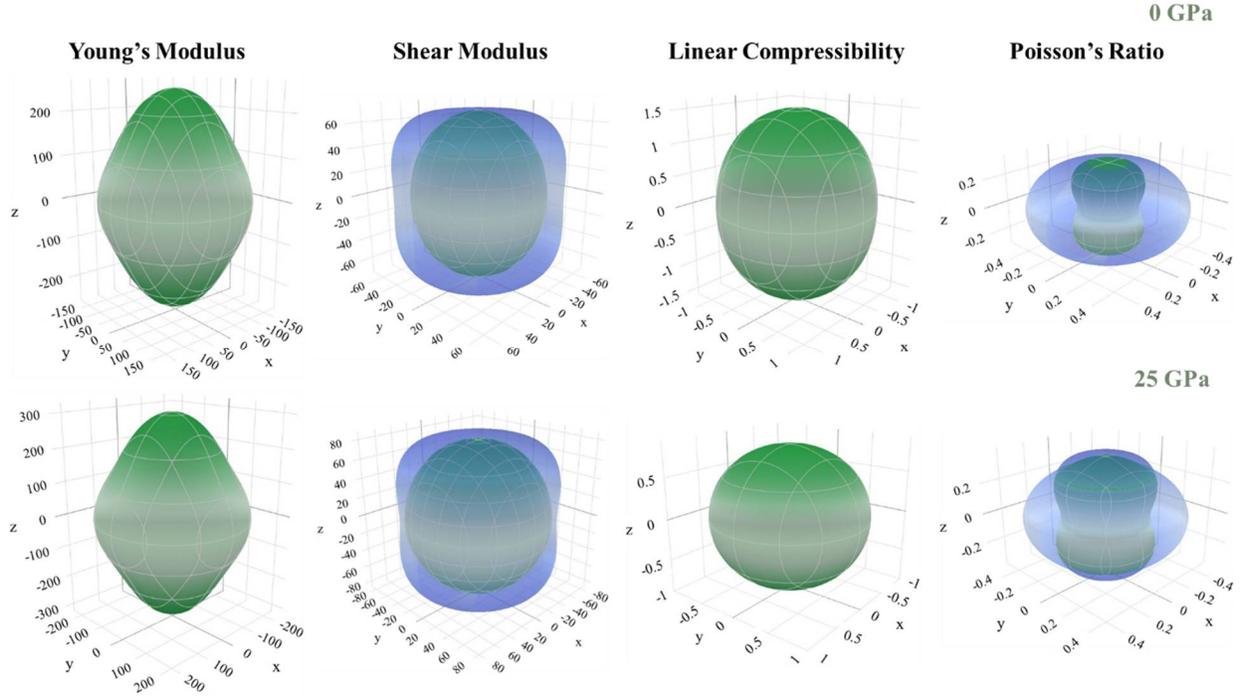

**Fig. 11.** 3D representations of the mechanical characteristics of ZrRe$_2$, showing Young's modulus, shear modulus, compressibility, and Poisson's ratio at 0 and 25 GPa. The transparent blue and green colors indicate the maximum and minimum points, respectively.

**Table 2:** The minimum and maximum values of Young's modulus (GPa), shear modulus (GPa), linear compressibility (TPa$^{-1}$), Poisson's ratio, and their ratios (anisotropy) for ZrRe$_2$ at 0 and 25 GPa.

| Pressure (GPa) | $E$ | | $A_E$ | $G$ | | $A_G$ | $\beta$ | | $A_\beta$ | $v$ | | $A_v$ |
|---|---|---|---|---|---|---|---|---|---|---|---|---|
| | $E_{min}$ | $E_{max}$ | | $G_{min}$ | $G_{max}$ | | $\beta_{min}$ | $\beta_{max}$ | | $v_{min}$ | $v_{max}$ | |
| 0 | 175.87 | 253.35 | 1.441 | 56.645 | 83.172 | 1.468 | 1.263 | 1.523 | 1.205 | 0.215 | 0.562 | 2.619 |
| 25 | 220.91 | 306.23 | 1.386 | 75.846 | 99.49 | 1.312 | 0.854 | 0.951 | 1.113 | 0.277 | 0.507 | 1.840 |

### 3.5. Thermo-physical properties

Over the past few years, theoretical studies on pressure-dependent thermal properties in metals have been widely conducted for high-pressure and high-temperature applications, as experimental measurements under such conditions are often challenging. The pressure dependent thermal behavior of ZrRe$_2$ was studied from Debye temperature, melting temperature, and phonon thermal conductivity. The formulae used to estimate these parameters are summarized elsewhere [29,41,42].

### 3.5.1. Debye temperature

Debye temperature ($\Theta_D$) is one of the most fundamental parameters reflecting many physical properties of solids, such as the bonding strength between elements, vibrational energy (thermal, acoustic, optical), specific heat capacity, melting temperature, indentation hardness, isothermal compressibility, and electron-phonon coupling constant. The estimated Debye temperature of ZrRe$_2$ at different pressures is displayed in Fig. 12a. Our calculated value of $\Theta_D$ at 0 GPa is slightly lower than the reference value. Since the Debye temperature of crystalline-solids depend both on the measurement method and temperature, we have employed a well-established model, namely the Anderson model. A gradual increase in $\Theta_D$ with applied pressure indicates thermodynamic stability of the material up to 25 GPa. Consequently, the covalent bonding strength and melting temperature are expected to increase with pressure as well. The moderate value of $\Theta_D$ suggests the material has potential for various applications, such as thermal barrier coatings (TBCs) and cryogenic electronics.

### 3.5.2. Melting temperature

Studying the melting temperature ($T_m$) of materials is crucial for various industrial applications, including manufacturing, electronics, aerospace, and energy. Fig. 12b shows the pressure-dependent variation of $T_m$ for ZrRe$_2$. At ambient pressure, ZrRe$_2$ exhibits a $T_m$ exceeding 2000 K, which increases with pressure—indicating excellent thermal stability, mechanical strength, and high resistance to wear. These properties suggest that ZrRe$_2$ is a promising candidate for thermal barrier coatings (TBCs) and applications under extreme conditions involving high pressure and temperature. Additionally, the $T_m$ vs. pressure plot provides valuable insights into the material's compressibility and potential phase transitions. The estimated $T_m$ value of ZrRe$_2$ at 0 GPa is lower than the experimental value reported in the literature. This discrepancy may be attributed to defect content in the experimental samples, whereas the melting temperature reported here corresponds to a perfect crystal.

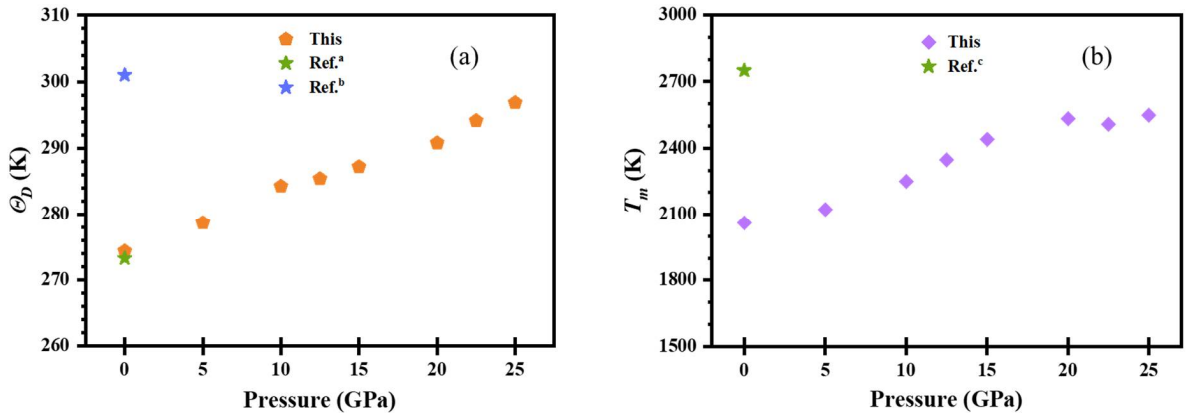

**Fig. 12.** Variation of (a) Debye temperature ($\Theta_D$) and (b) melting temperature ($T_m$) of ZrRe$_2$ with pressure. The reference data are shown as solid stars. Ref.[a] [15], Ref.[b] [17], and Ref.[c] [26].

*3.5.3. Phonon thermal conductivity*

The phonon thermal conductivity ($k_{ph}$) of engineering materials is a key factor in various technological applications, such as thermal barrier coatings (TBCs), thermal management in electronics, and thermoelectrics. High thermal conductivity materials are a primary focus in engineering, structural, micro- and nano-electronic devices. Conversely, materials with low thermal conductivity form the basis of a new generation of thermoelectric devices and TBCs. The $k_{ph}$ of a material is closely related to its bonding stiffness and the atomic masses of its constituent elements. The $k_{ph}$ of ZrRe$_2$ as a function of pressure and temperature is illustrated in Fig. 13. It's relatively low value of 2.65 W/m·K at 0 GPa and 300 K highlights the material's potential for applications in thermal coatings and thermoelectrics. Notably, $k_{ph}$ exhibits minimal pressure dependence up to 25 GPa, suggesting that its thermal conductivity remains stable under varying pressure conditions. Such pressure invariance may result from competing effects between the reduction in phonon lifetime and the increase in phonon group velocity under pressure [43]. Furthermore, the observed decrease in $k_{ph}$ with increasing temperature indicates enhanced phonon–phonon scattering due to temperature induced phonon anharmonicity.

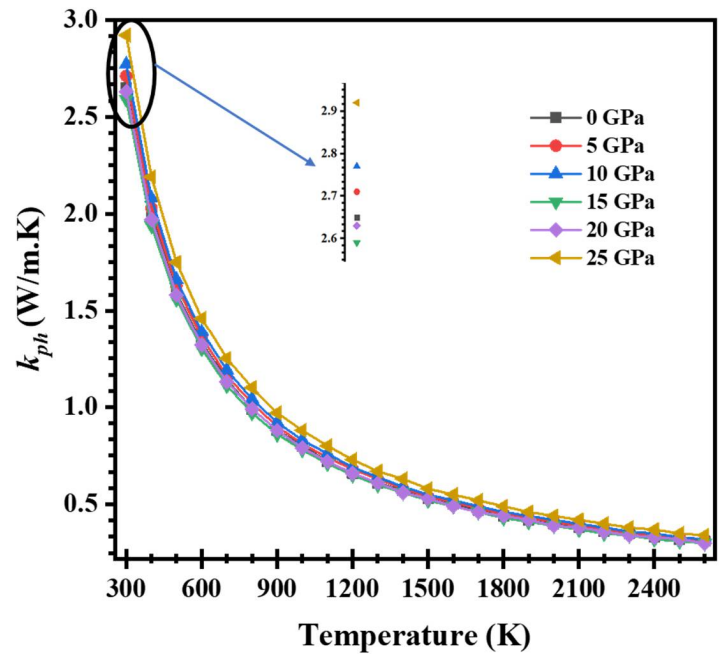

**Fig. 13.** Pressure- and temperature-dependent phonon thermal conductivity of ZrRe$_2$.

*3.5.4. Superconducting properties*

The superconductivity in Kagome intermetallics have recently attracted significant interest in condensed matter physics research community due to their unconventional superconducting properties [44,45]. W.L. McMillan [46] proposed a formula for calculating the transition temperature ($T_c$) of weakly to moderately strong coupled superconductors with a high degree of accuracy. The pressure-dependent $T_c$ of ZrRe$_2$ is estimated using McMillan's formula [46,47]:

$$T_c = \frac{\Theta_D}{1.45} \exp\left\{-\frac{1.04(1 + \lambda_{ep})}{\lambda_{ep} - \mu^*(1 + 0.62\lambda_{ep})}\right\} \qquad (4)$$

here, $\lambda_{ep}$ is the electron-phonon coupling constant, $\mu^*$ is the repulsive Coulomb pseudopotential, and $\Theta_D$ is the Debye temperature. The electron-phonon interaction energy ($V_{e-ph}$) is responsible for Fermi surface instability and Cooper pairing [48,49]. The electron-phonon coupling constant can be expressed as: $\lambda_{ep} = N(E_F)V_{e-ph}$. As the phonon spectrum of ZeRe$_2$ shows a weak pressure variation, it is reasonable to assume that the strength of $V_{e-ph}$ does not vary significantly with pressure. With this assumption, we can calculate the pressure dependence of $T_c$ of ZrRe$_2$ by taking into account of the pressure dependent variation of the parameters $\Theta_D$ and $N(E_F)$. At zero pressure, $\lambda_{ep}$ and $T_c$ of ZrRe$_2$ are found to be 0.69 and 6.1 K, respectively [17]. The pressure dependent $T_c$ calculated with SOC are depicted in Fig. 14. A typical value of $\mu^* = 0.13$ is used in our calculations [46]. The estimated value of bulk $T_c$ of ZrRe$_2$ at ambient pressure is found to be 5.8 K, in good agreement with the reported literature [17]. The value of $\lambda_{ep}$ indicates that ZrRe$_2$ is a moderately coupled superconductor, and an overall decreasing trend of $T_c$ with increasing pressure is observed. The inclusion of SOC slightly increases the $T_c$ value. This change is due to the SOC related changes in the electronic energy density of states at the Fermi level.

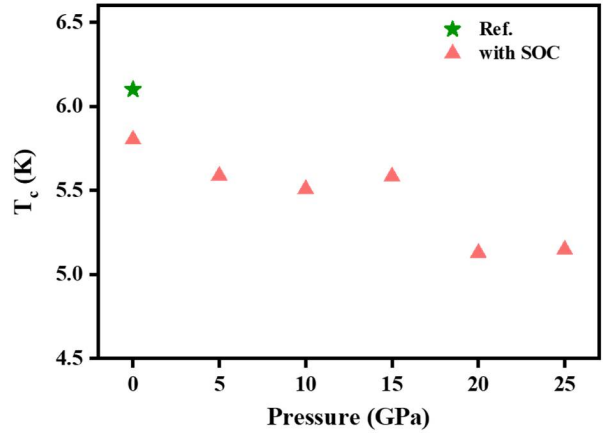

**Fig. 14:** The pressure-dependent superconducting transition temperature of ZrRe$_2$. The star symbol represents the reference value [17].

### 3.6. Phonon dynamic properties

Assessing the dynamical stability of materials is imperative for device applications, given that uniform mechanical stresses over time are unrealistic in real-world conditions. The calculated phonon dispersion (PD) and phonon density of states (PHDOS) of ZrRe$_2$ at 0 and 25 GPa without SOC are depicted in Fig. 15. The effect of SOC (not shown) is minimal. The absence of any negative (imaginary) frequency confirms the dynamical stability of the compound at both pressures. A material containing $N$ atoms per unit cell possesses 3 acoustic and ($3N$-3) optical vibrational modes. Accordingly, ZrRe$_2$ possesses a total of 36 vibrational modes, of which 3 are acoustic and 33 are optical modes. Among the optical modes, 8 are infrared (IR)-active, 12 are Raman-active, and 13 are silent. We have evaluated the optical phonon modes in ZrRe$_2$ at the $\Gamma$-point by the irreducible presentation of the point group ($D_6h$):

$$\Gamma(D^6h) = 3E_{2g} + 2E_{1g} + 3E_{1u} + 2A_{2u} + 2A_{1g}$$

where, *E* and *A* modes are doubly and singly degenerate, respectively. Table 3 lists the calculated frequencies of these active zone-center phonon modes of the compound at 0 GPa and 25 GPa. The obtained phonon spectra provide valuable reference data for future experimental investigations.

Acoustic phonons, caused by the coherent vibrations of atoms at sound velocity in a lattice outside their equilibrium position, are responsible for sound and heat conduction in solid materials [2,30]. The phonon spectrum of ZrRe$_2$ at 0 GPa spans up to about 6.85 THz, while extended by about 1.88 THz at 25 GPa. The acoustic phonon modes of ZrRe$_2$ get more linear with higher frequencies indicates that the thermal conductivity would increase at 25 GPa. This observation is consistent with our previously reported thermophysical studies (see section 3.5). The overlap (hybridization) between the acoustic and lower optical modes, along with the absence of an optical band gap, suggests that the compound exhibits no elastic and electromagnetic impedance, respectively. Such overlap enhances phonon scattering processes and consequently shortens the phonon lifetime, which can reduce lattice thermal conductivity. Absence of acoustic-optic phonon gap is due to the smaller atomic mass ratio. Typically, the small group velocities of optical phonons limit their direct contribution to thermal transport process, however, they play a crucial role by providing scattering channels for heat-carrying acoustic phonons. The heat capacity of a material is proportional to the PHDOS [50]. Peaks in PHDOS indicate low phonon dispersion. Partial PHDOS of ZrRe$_2$ confirms that the acoustic and lower optical modes are dominated by vibration of both Re1 atoms. The high frequency vibrations are associated mainly with Zr atoms. The contribution of Re1 is more significant than Re2 in the total PHDOS. The phonon frequencies increase and exhibit irregular changes under pressure, indicating bond strengthening within the compound.

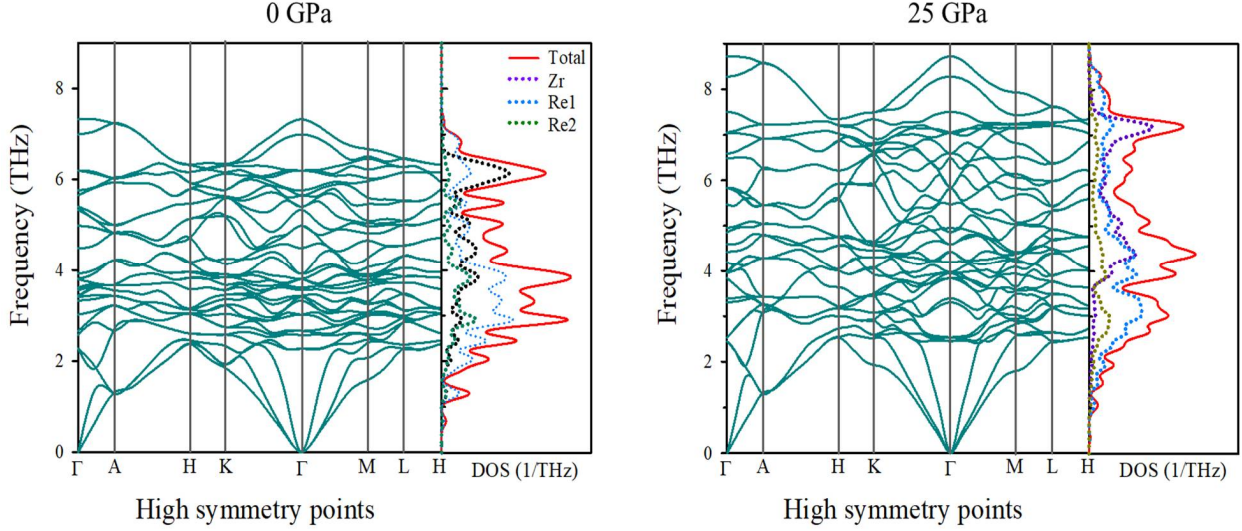

**Fig. 15.** Phonon dispersion curves with phonon density of states of ZrRe₂ at 0 and 25 GPa pressure without SOC.

**Table 3:** Calculated values of Raman and infrared (IR) active optical modes at the Γ-point of ZrRe₂ at 0 GPa and 25 GPa.

| Symmetry mode | Activity (cm$^{-1}$) | | | |
|---|---|---|---|---|
| | Raman | | IR | |
| | 0 GPa | 25 GPa | 0 GPa | 25 GPa |
| $E_{2g}$ | 75.658 | 81.795 | - | - |
| $E_{1g}$ | 100.531 | 127.910 | - | - |
| $E_{1u}$ | - | - | 113.414 | 131.918 |
| $A_{2u}$ | - | - | 118.472 | 165.353 |
| $E_{2g}$ | 126.595 | 147.772 | - | - |
| $E_{1u}$ | - | - | 130.557 | 153.107 |
| $A_{1g}$ | 177.914 | 216.574 | - | - |
| $E_{1g}$ | 180.276 | 194.614 | - | - |
| $E_{1u}$ | - | - | 188.557 | 219.730 |
| $E_{2g}$ | 206.783 | 234.796 | - | - |
| $A_{2u}$ | - | - | 206.875 | 250.335 |
| $A_{1g}$ | 233.201 | 276.136 | - | - |

## 3.7. Optical properties

The pressure-dependent optical properties of ZrRe₂ are derived from the complex dielectric function as a function of angular frequency ($\omega$): $\varepsilon(\omega) = \varepsilon_1(\omega) + i\varepsilon_2(\omega)$, where $\varepsilon_1(\omega)$ and $\varepsilon_2(\omega)$ represent the real (dispersive) and imaginary (absorptive) parts, respectively. The real and imaginary parts of the dielectric function, refractive index ($n$), extinction coefficient ($k$), absorption ($\alpha$), conductivity ($\sigma$), reflectivity ($R$), and loss function ($L$) of ZrRe$_2$ for [100] and [001] polarization directions of the electric field under 0 and 25 GPa are displayed in Fig. 16. A strong negative value at low energy in $\varepsilon_1(\omega)$ which diverges at $\omega \to 0$ highlights its metallic behavior. This strongly supports the findings obtained in the electronic properties (see section 3.2) and subsequent absorption and conductivity and studies (see Fig. 16e and 16f). A sharp negative value of $\varepsilon_1(\omega)$ at low energy is also a signatue of plasma oscillation. Consistent with the band structure analysis, this negative trend further supports the presence of Drude-like behavior. The calculated $\varepsilon_1(\omega)$ also describes the electronic polarizability of materials. Characteristic peaks of $\varepsilon_1(\omega)$ for [100] polarization are observed at 1.03 eV and 1.37 eV for 0 GPa and 25 GPa, respectively, in the IR region (Fig. 16a). In contrast, in [100] direction, the maxima of $\varepsilon_1(\omega)$ occurs at 2.94 eV and 1.00 eV for 0 GPa and 25 GPa, respectively. It is worth noting that the $\varepsilon_1(\omega)$ reaches zero from above at around 25 eV in the ultraviolet energy region, which indicates that ZrRe$_2$ becomes fairly transparent above 25 eV where optical absorption falls drastically. The $\varepsilon_1(\omega)$ spectra exhibit dependence on both polarization direction and pressure, up to about 6.00 eV. The plasma frequency of the compound, can also be estimated via zero crossing of the dielectric permittivity $\varepsilon_1(\omega)$, is 24.2 eV.

The imaginary part $\varepsilon_2(\omega)$ describing gain ($\varepsilon_2 > 0$) or absorption ($\varepsilon_2 < 0$) of photon. Large positive value of $\varepsilon_2(0)$ also reflects metallic nature of the material. The smaller the $\varepsilon_1(\omega)$ and the larger the $\varepsilon_2(\omega)$, the stronger the ability of the material to dissipate electromagnetic wave. Both intraband and interband transitions contributes to the imaginary part of the dielectric constants, $\varepsilon_2(\omega)$. Intraband transition, which dominate in metals, are primarily responsible for the low-energy features, particularly in the 0–0.3 eV energy range. Both $\varepsilon_1(\omega)$ and $\varepsilon_2(\omega)$ approach zero around 25 eV, a crucial point where several other optical parameters converge. At this energy, generally absorption coefficient (see Fig. 16e) and reflectivity (see Fig. 16g) drops sharply, and the energy loss function (see Fig. 16h) shows a notable peak [30,31,39].

The complex refractive index is defined as $N(\omega) = n(\omega) + ik(\omega)$, where $n(\omega)$ is refractive index and $k(\omega)$ is the extinction coefficient. The real part $n(\omega)$ describes the phase velocity of the photons in the medium, while the imaginary part $k(\omega)$ represents the energy loss of the electromagnetic wave in the medium, which is directly correlated to the dielectric function and absorption coefficient. The variation of $n(\omega)$ and $k(\omega)$ as a function of photon energy from 0–30 eV at 0 and 25 GPa are illustrated in Fig. 16(c–d). Both $n(\omega)$ and $k(\omega)$ attain their maximum values at low energies and decrease with increasing energy. This trend is attributed to the decreasing density of accessible electronic states away from the $E_F$. A distinct peak in $n(\omega)$ appears in the infrared (IR) region, indicating the material's potential for IR detection applications. In this region, $n(\omega)$ exhibits anisotropy and pressure dependence, becoming nearly isotropic at higher energies. The refractive index of ZrRe$_2$ remains relatively high across the IR to visible spectrum, suggesting its applicability in the design of optical and optoelectronic display devices.

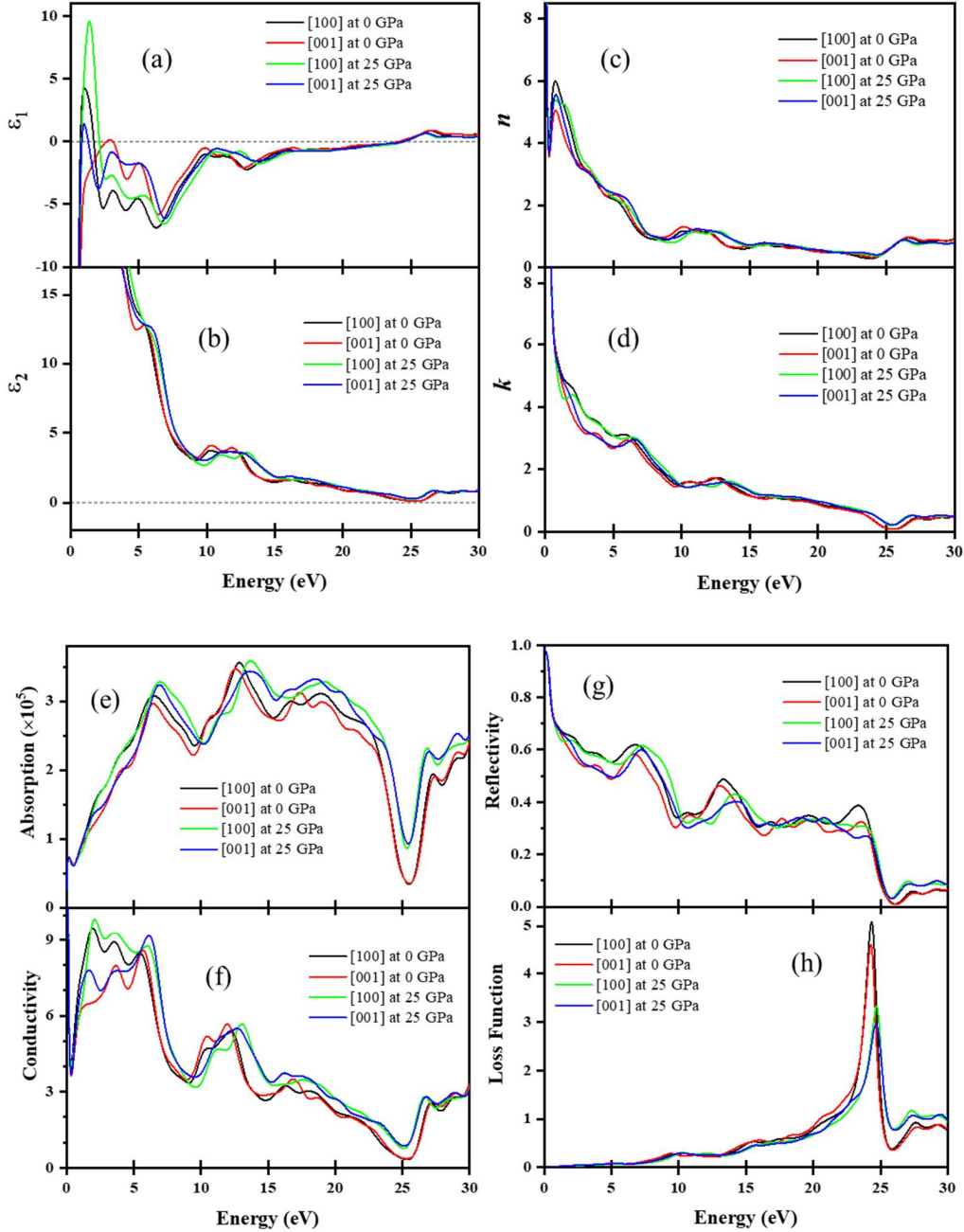

**Fig. 16.** The optical behavior of ZrRe$_2$ with energy along the [100] and [001] polarization directions, encompassing (a) real ($\varepsilon_1$) and (b) imaginary ($\varepsilon_2$) parts of the dielectric function, (c) refractive index ($n$), (d) extinction coefficient ($k$), (e) absorption coefficient, (f) optical conductivity, (g) reflectivity, and (h) loss function.

Fig. 16(e) illustrates the absorption spctra (α) of ZrRe$_2$ as a function of photon energy in [100] and [001] polarization directions at 0 GPa and 25 GPa. The absorption starts from zero photon energy for both polarizations, indicating metallic nature of the compound, which is consistent with the electronic structure analysis (see section 3.2). Overall, the absorption is higher along the [100] direction. The highest absorption peaks appear at approximately 12.87 eV and 12.55 eV for the [100] and [001] directions, respectively. The absorption remains pressure-insensitivit up to 6 eV, beyond which it shifts toward higher photon energies. This pressure stability response up to 6 eV suggests potential applicability in high-pressure and harsh-environment optical applications. A drastic drop in absorption is oserved at arround 22.5 eV.

Fig. 16f illustrates the photoconductivity $\sigma(\omega)$ at 0 and 25 GPa. $\sigma(\omega)$ starts rising from zero energy. This indicates absence of energy band gap of ZrRe$_2$ confirming its metallic nature. This is also consistent with the electronic band structure, optical absorption, and dielectric function studies.

The optical reflectivity $R(\omega)$ of a material defines the ratio of the reflected to incident photon energy. The static reflectivity $R(0)$ is about 99% which is also maximum within the studied energy range (see Fig. 16g) which predicts metallic and shiny nature of the material. The reflectivity spectra of ZrRe$_2$ are found to be maximum at the infrared region and stays above 52% for entire visible spectrum which is quite high compared to other intemetallic Kagome compounds [2,51,52]. The value of $R(\omega)$ remains above ~40% up to 8.9 eV. All these findings imply that ZrRe$_2$ can efficiently serve as a coating material in the infrared, visible, and UV energy regions. The reflectivity spectrum falls at three different regions (~0 eV, ~7 eV, and ~23 eV). The spectral shape reveals the strong damping behavior of charge carriers, suggesting that the charge carriers experience strong scattering in these energy regions. Additionally, several weak peak-like features appear in R(ω) above 10 eV. The reflectivity spectrum shows insignificant optical anisotropy and pressure dependence as well.

The energy loss function $L(\omega)$ of ZrRe$_2$ along [100] and [001] polarization directions at 0 and 25 GPa is displayed in Fig. 16h. $L(\omega)$ describes the energy loss of a fast electron traversing in a material. The peaks in $L(\omega)$ represents collective oscillation of charge carriers, a characteristics associated with plasmon resonance and the coresponding frequency is called plasma frequency ($\omega_p$). The $\omega_p$ can be expressed as: $\omega_p = \sqrt{(n_e e^2)/\varepsilon_0 m_e^*}$, where $n_e$ is the electron density, $e$ is the elementary charge, $\varepsilon_0$ is the permittivity of free space, and $m_e^*$ is the effective mass of the electrons. The position of $L(\omega)$ peak indicates the point of metallic-to-dielectric transition. In addition, the peaks of $L(\omega)$ also correspond to the trailing edges in the reflection and absorption spectra (see Fig. 16e and 16g).

## 4. Conclusions

In summary, a detailed investigation on the pressure dependent physical properties of intermetallic Kagome ZrRe$_2$ compound with and without SOC up to 25 GPa is presented. The structural properties reveal that the material is stable up to 25 GPa. The electronic studies confirmed metallic nature. The material possesses SOC effect and is nonmagnetic. All the three significant features (Dirac point, vHS, and flat band) of Kagome materials are reported in the electronic band structure of ZrRe$_2$ for the first time. The Fermi surface exhibits nesting tendency. The charge density difference plot indicates both covalent and ionic bonding. The calculated elastic constants and phonon dispersion ensure both mechanical and dynamical stability of ZrRe$_2$ within 0-25 GPa pressure range. ZrRe$_2$ is found to exhibit ductility and high machinability, making it an attractive material with promising application potential. The material exhibits moderate level of mechanical anisotropy that decreases under pressure. The compound is highly machinable with an excellent dry lubricating prospect and its performance improves under pressure, consistent with our previously reported results for intermetallic Kagome $CsV_3Sb_5$ [2]. The mechanical and thermo-physical properties indicate the material is promising as thermal barrier coatings (TBCs) and cryogenic electronics. The bulk superconducting transition temperature of $ZrRe_2$ decreases with increasing pressure. The optical properties of the compound confirm its metallic behavior. Moreover, the compound can serve as an efficient reflector of solar radiation in the infrared, visible, and UV energy regions. $ZrRe_2$ exhibits moderate level optical anisotropy.

In conclusion, we have systematically investigated a wide range of physical properties and their pressure dependence of the intermetallic Kagome compound $ZrRe_2$. The studied material exhibits promising potential for practical/engineering applications, for instance, in advanced electronic devices, spintronic systems, topological materials platforms, and thermal barrier coating–related applications. Therefore, we are optimistic that these findings will encourage further theoretical and experimental investigations on this compound.

**Acknowledgements**
Authors are grateful to the MPS Department, BRAC University, Dhaka-1212, Bangladesh, for providing the computing facilities for this work.

**Data availability**
The data sets generated and/or analyzed in this study are available from the corresponding author on reasonable request.

**CRediT authorship contribution statement**
**M. I. Naher:** Conceptualization, Software, Methodology, Formal analysis, Data curation, Visualization, Writing − original draft, review & editing; **A. F. M. Y. Haider:** Writing – review & editing, Validation, Supervision; **D. K. Paul:** Methodology, Data curation; **M. L. Rahman:** Writing –review & editing, Validation; **F. H. Haque:** Writing –review & editing, Validation; **S. H. Naqib:** Conceptualization, Formal analysis, Writing – review & editing, Validation, Supervision.

**Competing interests**
The authors declare that they have no known competing financial interests or personal relationships that could have appeared to influence the work reported in this paper.